\input amstex
\input amsppt.sty

\TagsOnRight \NoBlackBoxes
 \NoRunningHeads

\define\Y{\Bbb Y}
\define\Z{\Bbb Z}
\define\C{\Bbb C}
\define\R{\Bbb R}

\define\de{\delta}
\define\La{\Lambda}
\define\la{\lambda}
\define\si{\sigma}
\define\th{\theta}
\define\om{\omega}

\define\wt{\widetilde}
\define\wh{\widehat}
\define\tht{\thetag}
\define\down{{\downarrow}}
\define\up{\uparrow}

\define\Prob{\operatorname{Prob}}

\define\const{\operatorname{const}}

\topmatter
\title Stochastic dynamics related to Plancherel measure on
partitions
\endtitle

\author
Alexei Borodin and Grigori Olshanski
\endauthor

\dedicatory Dedicated to A.~M.~Vershik on the occasion of his
70th birthday \enddedicatory

\abstract Consider the standard Poisson process in the first
quadrant of the Euclidean plane, and for any point $(u,v)$ of
this quadrant take the Young diagram obtained by applying the
Robinson--Schensted correspondence to the intersection of the
Poisson point configuration with the rectangle with vertices
$(0,0)$, $(u,0)$, $(u,v)$, $(0,v)$. It is known that the
distribution of the random Young diagram thus obtained is the
poissonized Plancherel measure with parameter $uv$.

We show that for $(u,v)$ moving along any southeast--directed
curve $C$ in the quadrant, these Young diagrams form a Markov
process $\La_C$ with continuous time. We also describe $\La_C$
in terms of jump rates.

Our main result is the computation of the dynamical correlation
functions of such Markov processes and their bulk and edge
scaling limits.

\endabstract

\endtopmatter

\document
\head Introduction
\endhead

For any $n=1,2,\dots$, consider a measure on the set of all
partitions of $n$ which assigns to a partition $\la$ the square
of the dimension of the corresponding irreducible representation
of the symmetric group $S(n)$, divided by $|S(n)|=n!$. The
classical Burnside formula implies that this is a probability
measure (the sum of all weights equals one). It is usually
referred to as the $n$th {\it Plancherel measure}.

As was independently shown by Logan--Shepp \cite{LS} and
Vershik--Kerov \cite{VK1}, \cite{VK3} \footnote{A different
proof was later found by Kerov, see \cite{IO}.}, for large $n$
the random partitions distributed according to the $n$th
Plancherel measure have a typical (limit) shape. More detailed
information about local behavior of the random partitions in
different regions of the limit shape was later obtained in
Baik--Deift--Johansson \cite{BDJ1}, \cite{BDJ2}, Okounkov
\cite{Ok}, Borodin--Okounkov--Olshanski \cite{BOO}, Johansson
\cite{Jo1} for the ``edge'' of the limit shape, and in
Borodin--Okounkov--Olshanski \cite{BOO} for the ``bulk'' of the
limit shape.

One key observation that allowed to perform such a detailed
analysis was that the mixture of the Plancherel measures with
different $n$'s by a Poisson distribution, the so--called {\it
poissonized} Plancherel measure, has a nice algebraic structure:
it defines a {\it determinantal point process}\footnote{This means
that its correlation functions can be written as minors of a
suitable matrix called {\it the correlation kernel}.} \cite{BOO},
\cite{Jo1}.

In this work we construct stationary Markov processes on the set
of all partitions which have the poissonized Plancherel measures
as their invariant distributions. We prove that for any finite
number of time moments, the corresponding joint distribution of
the same number of random partitions defines a determinantal
point process, and we compute its correlation kernel.

As in the ``static'' case, there are two limit transitions, at
the edge and in the bulk of the limit shape. The corresponding
limit of the correlation kernel at the edge turns out to be the
well--known {\it extended Airy kernel}, while the limiting
kernel in the bulk appears to be a new one.

As a matter of fact, we obtain these results for more general,
{\it nonstationary} Markov processes on partitions, which change
the value of the poissonization parameter of the poissonized
Plancherel measure with time. This allows us to interpret the
results in terms of the Poisson process in a quadrant and its
projection by the Robinson--Schensted correspondence.

Our results also extend to more general measures on partitions,
the so--called {\it z--measures} \cite{BO1}. Markov processes
related to these measures are studied in detail in our paper
\cite{BO2}. The Plancherel measures may be viewed as appropriate
limits of the z--measures, and we use this connection
extensively in our proofs: the results of \S3 are obtained from
the similar results of \cite{BO2} by a degeneration.

Our work was largely inspired by previous papers due to
Okounkov--Reshetikhin \cite{OR}, Pr\"ahofer--Spohn \cite{PS},
and Johansson \cite{Jo2}. Thus, some of the results below may
already be known to experts. In particular, the results of
section 3 can be obtained using the formalism of Schur processes
\cite{OR} (this is not true for the z--measures, however), and
for certain special cases of our Markov processes the
determinantal structure of the correlation functions and the
edge scaling limit were obtained by Pr\"ahofer--Spohn \cite{PS}
in their work on polynuclear growth processes.

\subhead Acknowledgements\endsubhead This research was partially
conducted during the period the first author (A.~B.) served as a
Clay Mathematics Institute Research Fellow. He was also
partially supported by the NSF grant DMS-0402047. The second
author (G.~O.) was supported by the CRDF grant RM1-2543-MO-03.

\head 1. Construction of Markov processes
\endhead

As in Macdonald \cite{Ma} we identify partitions and Young
diagrams. By $\Y_n$ we denote the set of partitions of a natural
number $n$, or equivalently, the set of Young diagrams with $n$
boxes. By $\Y$ we denote the set of all Young diagrams, that is,
the disjoint union of the finite sets $\Y_n$, where
$n=0,1,2,\dots$ (by convention, $\Y_0$ consists of a single
element, the empty diagram $\varnothing$). Given $\la\in\Y$, let
$|\la|$ denote the number of boxes of $\la$ (so that
$\la\in\Y_{|\la|}$), and let $\ell(\la)$ be the number of nonzero
rows in $\la$ (the length of the partition).

For two Young diagrams $\la$ and $\mu$ we write $\mu\nearrow\la$
(equivalently, $\la\searrow \mu$) if $\mu\subset \la$ and
$|\mu|=|\la|-1$, or, in other words, $\mu$ is obtained from $\la$
by removing one box.

Let $\dim\la$, the {\it dimension\/} of $\la$, be the number of
all standard tableaux of shape $\la$. Equivalently, $\dim\la$ is
the dimension of the irreducible representation of the symmetric
group $S(|\la|)$ labelled by $\la$. A convenient explicit
formula for $\dim\la$ is
$$
\dim\la=\frac{n!}{\prod_{i=1}^N(\la_i+N-i)!}\,\prod_{1\le i<j\le
N}(\lambda_i-i-\la_j+j),\qquad \la\in\Y_n,
$$
where $N$ is an arbitrary integer $\ge \ell(\la)$ (the above
expression is stable in $N$).

For $\la\in\Y_n$, $\mu\in\Y_{n-1}$ set
$$
p^\down(n,\la;n-1,\mu)=\cases \dfrac{\dim\mu}{\dim\la},&\mu\nearrow\la,\\
0,&{\text{otherwise}},
\endcases
$$
and for $\la\in\Y_n$, $\nu\in\Y_{n+1}$ set
$$
p^\up(n,\la;n+1,\nu)=\cases \dfrac{\dim\nu}{\dim\la\,(n+1)},&\la\nearrow\nu,\\
0,&{\text{otherwise}}.
\endcases
$$
Then we have (see Vershik--Kerov \cite{VK2})
$$
\sum_{\mu\in\Y_{n-1}}p^\down(n,\la;n-1,\mu)=1, \qquad
\sum_{\nu\in\Y_{n+1}}p^\up(n,\la;n+1,\nu)=1.
$$

The $n$th {\it Plancherel measure\/} is a probability measure on
the finite set $\Y_n$ which is defined by
$$
M^{(n)}(\la)=\frac{(\dim\la)^2}{n!}\,, \qquad \la\in\Y_n\,, \tag
1.1
$$
see \cite{VK2}. The Plancherel measures with various indices $n$
are related to each other by means of the ``down probabilities''
$p^\down(n,\la;n-1,\mu)$ and the ``up probabilities''
$p^\up(n,\la;n+1,\nu)$, as follows (see \cite{VK2})
$$
\gathered
M^{(n-1)}(\mu)=\sum_{\la\in\Y_n}M^{(n)}(\la)p^\down(n,\la;n-1,\mu),\\
M^{(n+1)}(\nu)=\sum_{\la\in\Y_n}M^{(n)}(\la)p^\up(n,\la;n+1,\nu).
\endgathered  \tag1.2
$$
Note also the following relation:
$$
M^{(n)}(\la)p^\up(n,\la;n+1,\nu)
=M^{(n+1)}(\nu)p^\down(n+1,\nu;n,\la). \tag1.3
$$

Consider the Poisson distribution on the set
$\Z_+=\{0,1,2,\dots\}$, with parameter $\th>0$:
$$
Poisson_\th(n)=e^{-\th}\,\frac{\th^n}{n!}\,, \qquad n\in\Z_+\,.
\tag1.4
$$
Mixing all measures $M^{(n)}$ together by means of the Poisson
distribution \tht{1.4} we obtain a probability measure on the set
$\Y$. We denote it by $M_\th$ and call it the {\it poissonized
Plancherel measure\/} with parameter $\th$:
$$
M_\th(\la)=e^{-\th}\,\th^n\,\left(\frac{\dim\la}{n!}\right)^2,
\qquad n=|\la|. \tag1.5
$$

We are going to define a stationary Markov process
$\La_\th=\La_\th(t)$ with discrete state space $\Y$ and
continuous time $t\in\R$, and such that $M_\th$ is an invariant
measure of $\La_\th$. Moreover, $\La_\th$ is reversible with
respect to $M_\th$.

The trajectories of $\La_\th$ are step functions (in other
words, piece--wise constant functions) $\La(t)$ of variable
$t\in\R$, with values in $\Y$. We say that a trajectory $\La(t)$
makes a {\it jump\/} at a moment $t$ if the left limit
$\La(t^-)=\lim_{t'\up t}\La(t')$ differs from the right limit
$\La(t^+)=\lim_{t'\down t}\La(t')$. We reserve the notation
$\La_\th(t)$ to denote the {\it random\/} trajectory.

\example{Definition 1.1 (jump rates of $\La_\th$)} For any
$t\in\R$ and any $\la\in\Y_n$ we have by definition: conditional
on $\La_\th(t^-)=\la$, the probability that $\La_\th(\,\cdot\,)$
makes a jump to a diagram $\mu\in\Y_{n-1}$ in the time interval
$[t,t+dt]$ is equal to $R^\down(n,\la;n-1,\mu)dt+o(dt)$, where
$$
R^\down(n,\la;n-1,\mu)=n\,p^\down(n,\la;n-1,\mu). \tag1.6
$$
Likewise, the (conditional) probability of jumping to a diagram
$\nu\in\Y_{n+1}$ in the time interval $[t,t+dt]$ is equal to
$R^\up(n,\la;n+1,\nu)dt+o(dt)$, where
$$
R^\up(n,\la;n+1,\nu)=\th\,p^\up(n,\la;n+1,\nu).\tag1.7
$$
Finally, any other jumps in $[t,t+dt]$ are excluded (with
probability $1-o(dt))$. For obvious reasons we refer to \tht{1.6}
and \tht{1.7} as to the {\it jump rates\/}.
\endexample

The knowledge of the jump rates makes it possible (in our
concrete case) to define uniquely a {\it transition function\/}
$$
P_{\La_\th}(t,\la; s,\kappa)=\Prob\{\La_\th(s)=\kappa\mid
\La_\th(t)=\la\}, \qquad s>t, \quad \la,\kappa\in\Y,
$$
which depends only on $s-t$. The poissonized Plancherel measure
is compatible with the transition function,
$$
\sum_{\la\in\Y}M_\th(\la)P_{\La_\th}(t,\la;
s,\kappa)=M_\th(\kappa),
$$
which allows us to define the Markov process in question.
Moreover we can define the process not only on a half--line
$[t_0,+\infty)\subset\R$ but on the whole real line, that is, we
can construct a probability measure on the set $\{\La(t)\}$ of
$\Y$--valued step functions $\La(t)$ defined for all $t\in\R$.
\footnote{Let us note that the relevant step functions have only
finitely many jumps on finite time intervals $[t,s]\subset\R$.
Markov processes with such a property (finitely many jumps in
finite time) are sometimes called {\it regular\/}.} Since the
transition function is translation invariant in time, the
process is {\it stationary\/} (that is, the above measure on the
set $\{\La(t)\}$ is invariant under shifts of time, $t\to
t+\const$). Finally, the process turns out to be {\it
reversible\/} (that is, the measure on $\{\La(t)\}$ is also
invariant under the time reversion $t\to-t$).

\example{Remark 1.2} The above definition of the Markov process
$\La_\th$ can be rephrased as follows. Introduce an auxiliary
Markov process $N_\th$: a birth--death process on
$\Z_+=\{0,1,2,\dots\}$, which can be defined by the ``down'' and
``up'' jump rates
$$
R^\down(n;n-1)=n, \qquad R^\up(n;n+1)\equiv\th. \tag1.8
$$
The key property of $N_\th$ is that it has the Poisson
distribution \tht{1.4} as the invariant measure. \footnote{Recall
that this distribution is precisely the ``mixing'' measure used in
the definition of the poissonized Plancherel measure $M_\th$.} The
birth--death process $N_\th$ governs the jumps of $\La_\th$: each
moment $N_\th(\,\cdot\,)$ jumps down, say, from $n$ to $n-1$, the
trajectory $\La_\th(\,\cdot\,)$ makes a jump from $\Y_n$ to
$\Y_{n-1}$ (and the target Young diagram $\mu\in\Y_{n-1}$ is
chosen according to the ``down'' probabilities
$p^\down(n,\la;n-1,\mu)$, where $\la$ stands for the preceding
state). Likewise, when $N_\th(\,\cdot\,)$ jumps up, say, from $n$
to $n+1$, the trajectory $\La_\th(\,\cdot\,)$ makes a jump from
$\la$ to a random diagram $\nu\in\Y_{n+1}$ chosen according to the
``up'' probabilities $p^\up(n,\la;n+1,\nu)$. The fact that $M_\th$
is the invariant measure is deduced from the fact that the Poisson
distribution is the invariant measure of $N_\th$ and from
relations \tht{1.2}. The reversibility property of $\La_\th$ is
deduced from the reversibility of $N_\th$\; \footnote{Any
birth--death process is reversible.} and relation \tht{1.3}.
\endexample

The above construction of the Markov process can be generalized.
The idea is to use more general birth--death processes, with
time--dependent jump rates.

Let $\R^2_{>0}$ denote the open first quadrant of the Euclidean
plane $\R^2$ with coordinates $u>0,v>0$. Consider a parameterized
curve $C=(u(t),v(t))$ in $\R^2_{>0}$ subject to the following
conditions: the functions $u(t)>0$, $v(t)>0$ are continuous and
piece--wise continuously differentiable; the curve is directed
southeast, that is, $\dot u(t)\ge0$, $\dot v(t)\le0$, and $\dot
u(t)$ and $\dot v(t)$ do not vanish simultaneously (here the dot
means derivative with respect to $t$, and $t$ is interpreted as
time). Such curves $C$ will be called {\it admissible}.

We modify formulas \tht{1.8} as follows
$$
R^\down(n;n-1;t)=-n\,\frac{\dot v(t)}{v(t)}, \qquad
R^\up(n;n+1;t)=\dot u(t)v(t). \tag1.9
$$
Note that if $C$ is the hyperbola $uv=\th$ parameterized by $t=\ln
u$ then \tht{1.9} reduces to \tht{1.8}. In the general case we set
$$
\th(t)=u(t)v(t). \tag1.10
$$
There exists a birth--death process determined by the jump rates
\tht{1.9}; we denote it by $N_C$. Let
$$
P_{N_C}(t,n;s,m) =\Prob\{N_C(s)=m\mid N_C(t)=n\} \tag1.11
$$
be the transition function of $N_C$ (here $s>t$ and
$n,m\in\Z_+$). The process $N_C$ is no longer stationary in
time, and instead of a single Poisson distribution with fixed
parameter $\th$ (see \tht{1.4}) we deal with the whole family of
such distributions indexed by the varying parameter $\th(t)$
given by \tht{1.10}. These distributions are compatible with the
transition function \tht{1.11}:
$$
\sum_{n\in\Z_+}e^{-\th(t)}\,\frac{(\th(t))^n}{n!}
P_{N_{C}}(t,n;s,m) =e^{-\th(s)}\,\frac{(\th(s))^m}{m!}\,, \qquad
s>t, \quad m\in\Z_+\,.
$$
Therefore, we can make the assumption that
$$
\Prob\{N_{C}(t)=n\} =e^{-\th(t)}\,\frac{(\th(t))^n}{n!}
$$
for any $t$.

Now we can construct a Markov process $\La_{C}$ with state space
$\Y$ precisely as in Remark 1.2. It should be added that we
assume that conditional on $N_{C}(t)=n$, the distribution of
$\La_{C}(t)$ coincides with the $n$th Plancherel measure. This
implies that
$$
\Prob\{\La_{C}(t)=\la\} =M_{\th(t)}(\la)
$$
for any $\la\in\Y$.

\example{Remark 1.3} One can show that the Markov process
obtained from $\La_{C}$ by the reversion of time $t\to-t$ has a
similar form, $\La_{\wh C}$, where the curve $\wh C$ is the
image of $C$ under the transposition of coordinate axes. That
is, $\wh C$ is given by
$$
\wh u(t)=v(-t), \quad \wh v(t)=u(-t).
$$
\endexample

\example{Remark 1.4} A reparametrization of the curve $C$ (which
leaves it admissible) leads simply to a reparametrization of
time in the corresponding Markov process $\La_C$. There is a
distinguished parametrization of $C$ which is unique within an
additive constant:
$$
t=\frac12\,(\ln u-\ln v)+\const, \qquad (u,v)\in C. \tag1.12
$$
We call $t$ the {\it interior time\/} along the curve. If $C$ is
a hyperbola $uv=\th$, then the interior time coincides with the
natural time in the stationary Markov process $\La_\th$.
\endexample

\example{Remark 1.5} There are two particular cases of
nonstationary Markov processes $\La_{C}$ which can be called the
{\it descending\/} and the {\it ascending\/} ones. By
definition, they are obtained when we take as $C$ a vertical or
horizontal line, respectively. These processes correspond to
pure death or pure birth processes, respectively. More
generally, one considers broken lines $C$ with alternating
vertical and horizontal segments. Note that any admissible curve
can be approximated by such broken lines, which suggests the
idea that a general Markov process $\La_C$ can be approximated,
in an appropriate sense, by processes with alternating
descending and ascending fragments.
\endexample

\head 2. Interpretation of Markov processes via Poisson process
in quadrant
\endhead

In this section we describe a nice interpretation of the Markov
processes $\La_C$. As in \S1, we are dealing with the quadrant
$\R^2_{>0}\subset\R^2$ with coordinates $u>0,v>0$.

\example{Definition 2.1} Let $\pi$ be an $n$--point
configuration in $\R^2_{>0}$ such that no two points lie on the
same vertical or horizontal line. We assign to $\pi$ a
permutation $\si$ in the symmetric group $S(n)$ and then a Young
diagram $\la\in\Y_n$\,, as follows. Let $u_1<\dots<u_n$ and
$v_1<\dots<v_n$ be the $u$-- and $v$--coordinates of the points
in $\pi$. By definition, the permutation $\si$ determines a
matching of these coordinates. That is, the points in $\pi$ are
of the form $(u_i, v_{\si(i)})$, where $i=1,\dots,n$. Next, to
obtain $\la$ we apply the Robinson--Schensted algorithm (RS for
short). Recall (see, e.g., Sagan's book \cite{Sa, \S3.3 and
\S3.8}) that RS establishes an explicit bijection between
permutations $\si\in S(n)$ and pairs $(\Cal P, \Cal Q)$ of
standard Young tableaux of the same shape $\la\in\Y_n$, and we
just take this diagram  $\la$.
\endexample

Consider the Poisson process $\Pi$ in the quadrant $\R^2_{>0}$
with constant density 1. We also denote by $\Pi$ the random
point configuration in $\R_{>0}^2$ produced by the process. We
can assume that no two points in $\Pi$ lie on the same vertical
or horizontal line, because this condition holds for almost all
configurations $\Pi$.

\example{Definition 2.2} To any point $(u,v)\in\R^2_{>0}$ we
assign a random permutation $\si_\Pi(x,v)$ and a random Young
diagram $\la_\Pi(u,v)$, both depending on a realization $\Pi$ of
the Poisson process, as follows. Let $\square(u,v)$ denote the
rectangle with vertices $(u,v)$, $(u,0)$, $(0,v)$, $(0,0)$, and
let $\Pi(u,v)=\Pi\cap\square(u,v)$ be the random point
configuration in this rectangle. Then we set $\pi=\Pi(u,v)$,
$n=|\pi|$, and apply Definition 2.1.
\endexample

The construction of Definition 2.2 is well known. It was widely
used in the literature since Hammersley's paper \cite{Ha}. Note
that for a fixed point $(u,v)$, the number $n=|\Pi(u,v)|$ has
Poisson distribution \tht{1.4} with parameter $\th=uv$, and that
$\la_\Pi(u,v)$ is distributed according to the poissonized
Plancherel measure $M_\th$. Now we let $(u,v)$ vary.

\proclaim{Theorem 2.3} Let $C$ be an admissible curve in the
sense of \S1 and let a point $(u,v)=(u(t),v(t))$ move along $C$.
Let, as above, $\Pi$ be the random Poisson point configuration
in $\R^2_{>0}$, and consider the $\Y$--valued stochastic process
$\wt\La_C$ with random trajectories $\la_\Pi(u(t),v(t))$, where
the random Young diagram $\la_\Pi(u(t),v(t))$ is afforded by
Definition 2.2.

The process $\wt\La_C$ is a Markov process, equivalent to the
Markov process $\La_C$ of\/ \S1.
\endproclaim

By Theorem 2.3, each Markov process $\La_{C}$ can be interpreted
as a certain projection of the Poisson process in the quadrant.
Actually, a more precise result holds. Assume that the curve $C$
satisfies the conditions
$$
\lim_{t\to-\infty}u(t)=0, \qquad \lim_{t\to+\infty}v(t)=0, \tag2.1
$$
and let $\Cal D\subset\R^2_{>0}$ denote the subgraph of $C$,
that is, the part of the quadrant which is below and on the left
of $C$. For instance, in the case of the stationary process
$\La_\th$, one has $\Cal D=\{(u,v)\in\R^2_{>0}\mid uv<\th\}$.

\proclaim{Theorem 2.4} In the situation of Theorem 2.3, assume
additionally that condition \tht{2.1} is satisfied.

Then the construction of Theorem 2.3 provides a measure space
isomorphism between the realizations of the Poisson process in
the domain $\Cal D$ and the trajectories of $\wt\La_{C}$.
\endproclaim

Clearly, any trajectory $\{\la_\Pi(u(t),v(t))\}_{t\in\R}$ of
$\wt\La_C$ depends only on the restriction $\Pi|_{\Cal D}$ of
the corresponding Poisson configuration $\Pi$ to the domain
$\Cal D$. It turns out that, conversely, $\Pi|_{\Cal D}$ can be
reconstructed from $\{\la_\Pi(u(t),v(t))\}_{t\in\R}$. This
implies the theorem.

\example{Remark 2.5} Let, as in Theorem 2.3,  a point
$(u,v)=(u(t),v(t))$ move along an admissible curve $C$ and
replace $\la_\Pi(u,v)$ by $\si_\Pi(u,v)$, see Definition 2.2.
Then we obtain a random process taking values in permutations
$\si\in S(n)$ with varying $n$. One can show that this is again
a Markov process. Clearly, it ``covers'' the Markov process
$\wt\La_C$ of Theorem 2.3 (the latter is a projection of the
former). On the other hand, Theorem 2.4 implies a somewhat
paradoxical claim that the projection $\si\mapsto\la$ given by
the algorithm RS defines a measure space isomorphism between the
trajectories of both processes.
\endexample

\example{Remark 2.6} In case the curve $C$ is a straight line
$u+v=\const$, the corresponding Markov process $\La_C=\wt\La_C$
was earlier described in very different terms by
Pr\"ahofer--Spohn \cite{PS}, see also Remarks 3.4 and 4.5.
\endexample

\example{Remark 2.7} The assumption that the curve $C$ goes in
southeast direction is crucial for the Markov property in
Theorem 2.3. This can be demonstrated on the following simple
example. Consider three points in the quadrant: $a=(1,1)$,
$b=(2,1)$, and $c=(2,2)$. Then, conditional on $\la_\Pi(b)$ is
the one--box diagram, the random diagrams $\la_\Pi(a)$ and
$\la_\Pi(c)$ are not independent. This shows that on the broken
line going northeast, from $a$ to $b$ to $c$, the Markov
property does not hold.
\endexample

\head 3. Dynamical correlation functions \endhead

Consider the lattice of (proper) half--integers
$$
\Z'=\Z+\tfrac12=\{\dots,-\tfrac52,-\tfrac32,-\tfrac12,
\,\tfrac12,\,\tfrac32,\,\tfrac52,\dots\}.
$$
We can write $\Z'=\Z'_-\cup\Z'_+$, where $\Z'_-$ consists of all
negative half--integers and $\Z'_+$ consists of all positive
half--integers. For any $\la\in\Y$ we set
$$
\Cal L(\la)=\{\la_i-i+\tfrac12\mid i=1,2,\dots\}\subset\Z'.
$$
For instance, $\Cal L(\varnothing)=\Z'_-$. The correspondence
$\la\mapsto\Cal L(\la)$ is a bijection between the Young
diagrams $\la$ and those (infinite) subsets $\Cal L\subset\Z'$
for which the symmetric difference $\Cal L\triangle\Z'_-$ is a
finite set with equally many points in $\Z'_+$ and $\Z'_-$.

We regard $\Cal L(\la)$ as a point configuration on the lattice
$\Z'$. Assume we are given a probability measure $M$ on $\Y$.
Then we can speak about the random diagram $\la$ and hence about
the random point configuration $\Cal L(\la)$. The $n$th {\it
correlation function\/} of $M$ is defined as follows
$$
\rho_n(x_1,\dots,x_n)=\Prob\{x_1,\dots,x_n\in\Cal L(\la)\},
$$
where $n=1,2,\dots$ and $x_1,\dots,x_n$ are {\it pairwise
distinct\/} points of $\Z'$. In other words, the correlation
functions tell us what is the probability that the random point
configuration $\Cal L(\la)$ contains a given finite set of
points. The collection of all correlation functions determines
the initial probability measure $M$ uniquely.

For the poissonized Plancherel measure $M_\th$ defined in
\tht{1.5} the correlation functions were found independently by
Borodin--Okounkov--Olshanski \cite{BOO} and Johansson \cite{Jo1}:

\proclaim{Theorem 3.1} {\rm(i)} The correlation functions of the
measure $M_\th$ have determinantal form
$$
\rho_n(x_1,\dots,x_n)=\det\limits_{1\le i,j\le n}[K(x_i,x_j)],
$$
where $n=1,2,\dots$ and
$$
K(x,y)=\sqrt\th\;\frac{J_{x-\tfrac12}(2\sqrt\th)
J_{y+\tfrac12}(2\sqrt\th)-J_{y-\tfrac12}(2\sqrt\th)
J_{x+\tfrac12}(2\sqrt\th)}{x-y}\,. \tag3.1
$$
Here $J_z(\,\cdot\,)$ denotes the Bessel function which is
regarded as a function of its index $z$. When $x=y$, the
indeterminacy in formula \tht{3.1} is resolved by l'H\^opital's
rule.

{\rm(ii)} The kernel can also be written in the form
$$
K(x,y)=\sum_{a\in\Z'_+}
J_{x+a}(2\sqrt\th)J_{y+a}(2\sqrt\th).\tag3.2
$$
\endproclaim

The function $K(x,y)$ on $\Z'\times\Z'$ is called the {\it
discrete Bessel kernel\/}.

We are now going to state an analog of Theorem 3.1 for the
Markov processes $\La_C$. The correspondence $\la\mapsto\Cal
L(\la)$ allows us to regard each trajectory $\La(t)$ in $\Y$ as
a varying in time  point configuration $\Cal L(\La(t))$. In this
picture, a jump consists in shifting one of the points of the
configuration by $\pm1$.

\example{Definition 3.2} Let $\La(t)$ stand for the random
trajectory of a $\Y$--valued stochastic process with time
parameter $t$. By the $n$th {\it dynamical correlation
function\/} of the process we mean the function
$$
\rho_n(t_1,x_1;\dots;t_n,x_n)=\Prob\{x_i\in\Cal L(\La(t_i)),
i=1,\dots,n\}
$$
where $(t_1,x_1), \dots,(t_n,x_n)$ are pairwise distinct
elements of $\R\times\Z'$.
\endexample

In other words, the dynamical correlation functions tell us what
are the probabilities of the following events: for each $t$ in a
given finite subset of $\R$, the point configuration $\Cal
L(\La(t))$ contains a given finite subset $X(t)\subset\Z'$. Note
that this definition somewhat resembles that of the
finite--dimensional distribution, because both definitions
describe the behavior of the process at a finite number of
moments of time.

\proclaim{Theorem 3.3} Let $C=(u(t),v(t))$ be an admissible
curve in $\R^2_{>0}$ with its canonical parametrization
\tht{1.12} given by interior time $t$, let $\th(t)=u(t)v(t)$ as
in \tht{1.10}, and let $\La_C$ be the corresponding Markov
process.

{\rm(i)} The dynamical correlation functions of $\La_C$ have
determinantal form
$$
\rho_n(t_1,x_1;\dots;t_n,x_n)=\det\limits_{1\le i,j\le
n}[K_C(t_i,x_i;t_j,x_j)],
$$
where $n=1,2,\dots$ and $K_C(s,x;t,y)$ is a kernel on
$(\R\times\Z')\times(\R\times\Z')$ which can be written as a
double contour integral
$$
K_C(s,x;t,y)=\frac{e^{\frac12(s-t)}}{(2\pi
i)^2}\int\limits_{\{\om_1\}}\int\limits_{\{\om_2\}}
\frac{e^{\sqrt{\th(s)}(\om_1-\om_1^{-1})+\sqrt{\th(t)}(\om_2-\om_2^{-
1})}}{e^{s-t}\om_1\om_2-1}
\om_1^{-x-\tfrac12}\om_2^{-y-\tfrac12}d\om_1\,d\om_2 \tag3.3
$$
where $\{\om_1\}$ and $\{\om_2\}$ are any two contours which go
around 0 in positive direction and satisfy the following
condition:

$\bullet$ if $s\ge t$, so that $e^{s-t}\ge1$, then the contour
$\{\om_1\}$ must contain the contour $\{e^{t-s}\om_2^{-1}\}$;

$\bullet$ if $s< t$, so that $e^{s-t}<1$, then, on the contrary,
the contour $\{\om_1\}$ must be contained in the contour
$\{e^{t-s}\om_2^{-1}\}$.

{\rm(ii)} The kernel can also be written in the form
$$
K_C(s,x;t,y)=\pm\sum_{a\in\Z'_+} e^{-a|s-t|}\,J_{x\pm
a}(2\sqrt{\th(s)})J_{y\pm a}(2\sqrt{\th(t)}) \tag3.4
$$
where the plus sign is chosen if $s\ge t$ whereas the minus sign
is chosen if $s<t$.
\endproclaim

One can verify that if $s=t$ then \tht{3.3} can be reduced to
\tht{3.1}, and it is immediate that \tht{3.4} turns into
\tht{3.2}.

Note the asymmetry between the conditions $s\ge t$ and $s<t$.
One can show that
$$
\lim_{s\to t^+} K_C(s,x;t;y) =K_C(t,x;t,y)=\lim_{s\to
t^-}K_C(s,x;t;y)+\de_{xy}
$$
This agrees with the fact that
$$
\sum_{a\in\Z'}
J_{x+a}(2\sqrt\th)J_{y+a}(2\sqrt\th)=\delta_{xy}\,.
$$
\example{Remark 3.4} For the curves $u+v=\const$,  formula
\tht{3.4}  was earlier proved by Pr\"ahofer and Spohn, \cite{PS,
\tht{3.52}}.
\endexample

\head 4. Scaling limits
\endhead

The study of the asymptotic behavior of the random Young
diagrams distributed according to the $n$th Plancherel \tht{1.1}
as $n\to\infty$, or according to the poissonized Plancherel
measure \tht{1.5} as $\th\to\infty$, is an important and
nontrivial problem which has many different aspects. We refer
the reader to \cite{BOO}, \cite{IO} for a general discussion and
relevant references, and restrict ourselves here to considering
two types of the asymptotics: in the middle (bulk) and at the
edge of the Young diagrams distributed according to $M_\th$ and
Markov processes introduced in \S1.

\subhead{Bulk}\endsubhead We start with recalling the following
statement.

\proclaim{Theorem 4.1 (\cite{BOO, Section 3})} The correlation
functions of the measure $M_\th$ have the following limit as
$\th\to\infty$: Fix $c\in(-2,2)$ and let $x_0(\th)\in\Z'$ be
such that
$$
x_0(\th)=c\cdot\sqrt\th +o(\sqrt\th),\quad \th\to\infty.
$$
Further, for any $n=1,2\dots$ and arbitrary $x_1,\dots,
x_n\in\Z$, set
$$
x_i(\th)=x_0(\th)+x_i, \qquad i=1,\dots,n.
$$
Then
$$
\lim_{\th\to\infty}\rho_n(x_1(\th),\dots,x_n(\th))=\det_{1\le
i,j\le n}[S_c(x_i-x_j)],
$$
where
$$
S_c(r)=\frac{\sin\bigl(\arccos(c/2)\cdot r\bigr)}{\pi r}\,. \tag
4.1
$$
\endproclaim

The function $S_c(x-y)$ on $\Z\times\Z$ is called the {\it
discrete sine kernel}.

In Theorem 4.1, the pairwise distances $x_i(\th)-x_j(\th)$ do
not vary as $\th\to+\infty$; if instead this one supposes that
for some $i$ and $j$, the distance between  two points
$x_i(\th)$ and $x_j(\th)$ from the bulk\footnote{meaning that
$x_i(\th)/\sqrt\th$ and $x_j(\th)/\sqrt\th$ are bounded away
from the edges $-2$ and $2$.} tends to infinity together with
$\th$, then the events of finding particles (=elements of $\Cal
L(\la)$) at these locations become asymptotically independent.

Our goal is to extend Theorem 4.1 to the Markov processes
$\La_{C}$.

Let us return for a moment to the interpretation of $\La_{C}$
via the Poisson process in the first quadrant, see \S2. Theorem
4.1 deals with Young diagrams $\la_\Pi(u,v)$ sitting at the
points $(u,v)$ such that $uv=\th\to\infty$. Now let us assume
that such a point $(u,v)$ sits on an admissible curve.

It turns out that the bulk of $\la_\Pi(u,v)$ will have nontrivial
correlations with the bulks of the Young diagrams corresponding to
other points of the curve if the distance to these points in the
$(u,v)$--plane remains finite as $\th\to\infty$. If we assume that
we are far enough from the boundary of the quadrant ($u/v$ is
bounded away from zero and infinity) then the interior time change
between such points has to be of order $1/\sqrt{\th}$.

\proclaim{Theorem 4.2} Let $C_\th=(u_\th(t),v_\th(t))$ be a
family of admissible curves in $\R^2_{>0}$ with their canonical
parametrizations given by the interior time \tht{1.12}. Here
$\th>0$ is a parameter, and we assume that there exists a
constant $T\in\R$ such that $u_\th(T)v_\th(T)=\th$. For
instance, we may take as the curves $C_\th$ the hyperbolas
$uv=\th$.

The dynamical correlation functions of $\La_{C_\th}$ have the
following limit as $\th\to+\infty$. Fix an arbitrary
$c\in(-2,2)$ and let $x_0(\th)\in\Z'$ be such that
$$
x_0(\th)=c\cdot\sqrt\th +o(\sqrt\th),\quad \th\to\infty.
$$
Let $n=1,2,\dots$, and let $\tau_1, \dots,\tau_n\in\R$ and
$x_1,\dots,x_n\in\Z$ be arbitrary. Further, assume that
$$
t_i(\th)=T+\tau_i/\sqrt{\th}+o(1/\sqrt{\th}).
$$
Then
$$
\lim_{\th\to\infty}\rho_n(t_1(\th),x_0(\th)+x_1;\dots;t_n(\th),
x_0(\th)+x_n) =\det\limits_{1\le i,j\le
n}\left[S_c(\tau_i-\tau_j;x_i-x_j)\right],
$$
where
$$
S_c(h;r)=\frac 1{2\pi i}\int\limits_{\{\om\}}
e^{-h(\omega+\omega^{-1}-c)}\frac{d\om}{\om^{r+1}}\,, \tag 4.2
$$
and $\{\om\}$ is a contour in $\C$ going from the point
$e^{-i\phi}$ to the point $e^{i\phi}$, $\phi=\arccos(c/2)$, in
such a way that it passes to the right of the origin if $h\ge
0$, and to the left of the origin if $h<0$.
\endproclaim

Note that the limit correlations do not depend on the choice of
the curves $C_\th$.

It is readily verified that when $h=0$, \tht{4.2} coincides with
\tht{4.1}.

Another, somewhat similar extension of the discrete sine kernel
called {\it incomplete beta kernel} was obtained by
Okounkov--Reshetikhin \cite{OR, Section 3}.

\subhead Edge \endsubhead We now concentrate our attention on the
asymptotics of the correlation functions at the edge of the Young
diagrams, where $x_i(\th)\sim \pm 2\sqrt{\th}$. Due to the
symmetry of our measures with respect to transposition of Young
diagrams, which also swaps the edges $2\sqrt{\th}$ and
$-2\sqrt{\th}$, it suffices to consider one of the edges.

\proclaim{Theorem 4.3 (\cite{BOO, Section 4}, \cite{Jo1})} The
correlation functions of $M_\th$ have the following scaling
limit as $\th\to\infty$: For any $n=1,2,\dots$ and any
$x_1,\dots,x_n\in\R$, let $x_1(\th),\dots,x_n(\th)\in\Z'$ be
such that
$$
x_i(\th)=2\sqrt{\th}+x_1\th^{\frac 16}+o\left(\th^{\frac
16}\right), \qquad i=1,\dots,n.
$$
Then
$$
\lim_{\th\to\infty}(\th^{\frac
16})^n\rho_n(x_1(\th),\dots,x_n(\th))=\det_{1\le i,j\le n}[\Cal
A(x_i,x_j)],
$$
where
$$
\Cal A(x,y)=\frac{Ai(x)Ai'(y)-Ai'(x)Ai(y)}{x-y}\,,
$$
and $Ai(x)$ is the Airy function.
\endproclaim

Note that the factor $(\th^{\frac 16})^n$ comes from the scaling
of the space variable $x$.

The function $\Cal A(x,y)$ on $\R^2$ is called the {\it Airy
kernel}. Another useful formula for the Airy kernel is
$$
\Cal A(x,y)=\int_{0}^\infty Ai(x+a)Ai(y+a)da.
$$

Once again, let us return to the interpretation of $\La_{C}$
through the Poisson process. It turns out that the edges of the
Young diagrams sitting on a curve $(u(t),v(t))$ with
$uv=\th\to\infty$ will have nontrivial correlations if the
distance between the points in the $(u,v)$--plane grows as
$\th^{\frac 13}$. This means that if we are away from the
coordinate axes then the interior time change is of order
$\th^{-\frac 16}$.

\proclaim{Theorem 4.4} Let $C_\th=(u_\th(t),v_\th(t))$ be a
family of admissible curves in $\R^2_{>0}$ with their canonical
parametrizations given by the interior time \tht{1.12}. Here
$\th>0$ is a parameter, and we assume that there exists a
constant $T\in\R$ such that $u_\th(T)v_\th(T)=\th$. For
instance, we may take as the curves $C_\th$ the hyperbolas
$uv=\th$.

The dynamical correlation functions of $\La_{C_\th}$ have the
following limit as $\th\to+\infty$. For arbitrary
$\{\tau_i\}_{i=1}^n\subset\R$, choose $t_i(\th)$ such that
$$
t_i(\th)=T+\tau_i\,{\th}^{-\frac 16}+o\left({\th}^{-\frac
16}\right), \qquad i=1,\dots,n.
$$
Further, for arbitrary $x_1,\dots,x_n\in\R$, choose
$x_1(\th),\dots,x_n(\th)\in\Z'$ such that
$$
x_i(\th)=2\sqrt{u_\th(t_i(\th))v_\th(t_i(\th))}+x_i\th^{\frac
16}+o\left(\th^{\frac 16}\right), \qquad i=1,\dots,n.
$$
Then
$$
\lim_{\th\to\infty}(\th^{\frac 16})^n\rho_n(t_1(\th),
x_1(\th)\,;\,\dots\,; \,t_n(\th), x_n(\th)) =\det\limits_{1\le
i,j\le n}\left[\Cal A(\tau_i-\tau_j;x_i,x_j)\right],
$$
where
$$
\Cal A(\tau;x,y)=\cases \int_0^\infty e^{-\tau
a}Ai(x+a)Ai(y+a)da,&\tau\ge
0,\\
-\int_0^{+\infty}e^{-|\tau| a}Ai(x-a)Ai(y-a)e^{-\tau
a}da,&\tau<0.
\endcases
$$
\endproclaim

The kernel $\Cal A(\tau;x,y)$ is called the {\it extended Airy
kernel}. It is stationary in time.

\example{Remark 4.5} A special case of Theorem 4.4 with $C_\th$
being the lines $u+v=\const$ was proved in Pr\"ahofer--Spohn
\cite{PS}.
\endexample

The next statement uses the notion of the {\it Airy process},
introduced in \cite{PS}, see also Johansson \cite{Jo2}.

\proclaim{Corollary 4.6} Let $\{(u_\th(t),v_\th(t)\mid
t\in(T-\epsilon,T+\epsilon)\}$ be a family of admissible curves
in $\R^2_{>0}$ with their canonical parametrizations, and
$u_\th(T)v_\th(T)=\th$. Denote by $l(t,\th)$ the length of the
first row of the random Young diagram
$\la_\Pi(u_\th(t),v_\th(t))$ and set $t(\tau)=T+\tau\th^{-\frac
16}$. Then as $\th\to\infty$, the random variable
$$
L(\tau)=\frac{l(t(\tau),\th)-
2\sqrt{u_\th(t(\tau))\,v_\th(t(\tau))}}{\th^{\frac 16}}
$$
converges, as a function of $\tau$, to the Airy process.

\endproclaim

\bigskip

{\smc A.~Borodin}: Mathematics 253-37, Caltech, Pasadena, CA
91125, U.S.A.,

\medskip

E-mail address: {\tt borodin\@caltech.edu}

\bigskip

{\smc G.~Olshanski}: Dobrushin Mathematics Laboratory, Institute
for Information Transmission Problems, Bolshoy Karetny 19,
127994 Moscow GSP-4, RUSSIA.

\medskip

E-mail address: {\tt olsh\@online.ru}

\enddocument
\end